\documentclass[a4paper,10pt]{article}
\usepackage[utf8x]{inputenc}
\usepackage{color}

\title{ Convection in  Drying and Freezing  Ground}
\author{Mir Faizal and Stephen Peppin \\ 
OCCAM, Mathematical Institute, University of Oxford
\\ Oxford
OX1 3LB, United Kingdom  }

\begin{document}

\maketitle

\begin{abstract}

In this paper we  analyse the drying of a  soil composed of particles, water and solute impurities, and study the occurrence of 
 convective instabilities during evaporation.  
 We find that the main 
driving force  for instability is  the formation of a concentration gradient  at the soil surface due to the evaporation of water. 
A similar  phenomenon may occur during the thawing of frozen ground in  Arctic regions.

\end{abstract}

\section{Introduction}

The formation of a surface seal in soil, 
the hardening  of the  seal to form a  crust and cracking of this crust, 
 has been observed to occur under the  combined influence of rain and dry weather \cite{1a, 2a}.  
The sealing leads to a local increase in the bulk density and a decrease in porosity along with a 
decrease in the hydraulic conductivity \cite{4a}. After drying these surface seals transform into 
surface crusts. The sealing and crusting of soil surfaces increases the runoff in the soil surfaces.
In fact, initially the soil loss reduces because soil strength increases during crusting. However, crusting 
eventually leads to the formation of cracks which increases runoff in the soil surfaces \cite{5a}. 
The  detailed effect of crusting on soil loss depends on various factors like the distribution 
of crusted and non-crusted regions and the shrinkage cracks within the crusted layer \cite{6a}. 
Cracking depends critically on the rate of drying and so it  has been observed to occur
within  a few days of a rainstorm in a dry season \cite{6b}. 

The effects of sealing and crusting on soil loss are most visible in agricultural lands. 
This is because the 
sealing and crusting is most likely to occur in the soil which is not vegetated in the agricultural environment.
In Europe  a quarter of its agricultural land
exhibits some form of soil erosion risk  \cite{7ab}.
 In fact, a 20 mm rain has been 
observed to form surface seal and cause soil loss in certain hilly areas \cite{7a,7b}. In 
this case, the thickness of the surface seal 
 was observed to be  on the order of a few millimeters. However, the formation of a 7 cm thick surface seal during a wet season has also 
been observed \cite{8a}.  
Since the soil loss can effect the properties of soil  and 
thus have a major impact  on agricultural production, it is important to understand the physical processes at work in  drying ground. 
 In this paper we analyse  convective instabilities that are observed to occur during the evaporation of a mixture of soil particles and water 
\cite{ee}. 

Evaporation of pure liquids can  cause convection in  those liquids \cite{a,b,c,d,e,f}. 
This is because  the  evaporation causes the surface temperature to drop and thus 
sets up  an unstable density gradient which can induce a Rayleigh–Bernard instability. 
In  addition, for most 
 liquids  surface tension is a function of
temperature and so perturbations in temperature along the film surface also create  perturbations in the
surface tension.  The liquid then flows from places of lower surface tension to places of
 higher surface tension. This  flow is called the Marangoni effect. 
Thus, both the temperature 
gradient and  surface tension variations contribute to the occurrence of convection  in pure liquids. 
 The combined phenomenon is called  Bernard–Marangoni convection.

The  formation of surface seals and crusts can be understood by analysing the drying 
of pools of muddy water. Evaporation in mixtures also  leads to convection \cite{aa,ba,ca,da}. 
In this case as the liquid drys up a concentration gradient is also set up along with a temperature gradient.
The liquid moves under the combined influence of both these gradients. Furthermore, the  Marangoni effect also 
occurs due to the dependence of the of surface tension on both the temperature and the concentration. Thus, 
a perturbation in either the temperature or the concentration  can  cause convection. 
Convection has been observed to occur in  the course of evaporation of a binary mixture of  soil particles and water \cite{ee}. 
In doing so a four weight percent mixture of bentonite and water was analysed at twenty  degrees centigrade. 
It was found that convection occurred  during the course of 
drying of this mixture and  produced hexagonal patterns. However, it was also observed that the 
temperature gradient set up during  the evaporation could not explain the occurrence of  convection in this 
 system. In this paper we will show that the concentration 
gradient is  the main driving force behind such convection during the drying of ground. 

 We also study the thawing of frozen ground which occurs in the Arctic circle.
 Freezing and thawing of  soil
in the Arctic circle results in the formation of various  surface patterns such as soil hummocks and stone circles \cite{ar1, ar2, br1, br2}.
 Various models have 
been proposed to explain the occurrence of these patterns. 
In one of these models the maximum density of water at four degree centigrade sets up the convection \cite{ar4, ar41, ar5}. 
This model is based on the  convection of water through soil pores which has not been observed for the soils under consideration. 
It has also been proposed that the  sedimentation of soil during thawing 
 sets up an unstable density profile which  causes convection \cite{ara1, ara2, ara0}. 
However, the measured soil density gradient is not large enough to initiate convection \cite{ara4}. 
Another model  proposes that the   motion responsible for pattern
 formation occurs during the freezing of ground (differential frost heave) \cite{ar01, ar02}.
 This model yields a plausible mechanism for patterned ground but has yet to be experimentally 
confirmed, and does not explain certain observations such as the soil convection observed to occur in later summer \cite{br1}.

The  soil convection can potentially be explained  by a  phenomenon similar to the one that 
occurs in the formation of surface seals. The ice melts in late summer and
at the same time  evaporation takes place from the surface of the  soil. This causes an unstable density profile to develop 
that is in principle large enough to initiate wholesale soil
 convection. 
This convection has been observed in fields and is thought to contribute to the patterns that form in Arctic region \cite{br1, ara0}.  
It may be noted that similar 
patterns have been detected  on Mars, suggesting that in the past the temperature on Mars 
 may have been large enough for the ice to  melt  and  soil convection to take place  \cite{mars}.

In this paper we will use tensor notations for performing  the  stability analysis  \cite{ti}. 
We will also use a new approach based on Green's functions to perform stability analysis. As will be evident 
that the stability analysis of a ternary mixture is  very complicated, and it would not be possible to 
perform it using the conventional methods that are usually employed for stability analysis. 

\section{Ternary Mixture}
In this section we will first analyse the evaporation  of a  mixture of  soil particles and water. 
As the liquid evaporates, both the temperature and the concentration  change at the surface. 
This causes an instability to occur which drives convection. In order to analyse the occurrence of this stability, we first 
observe that the density of the liquid depends on both the temperature of this  mixture
and concentration of  particles in water.
However, in any real situation there will also be various  solute impurities in the soil. 
Hence, we  analyze a ternary mixture of water, soil particles and dissolved solutes. Thus, we can write, 
\begin{equation}
 \rho = \rho_0 [ 1 - \alpha_c (C- C_0) + \alpha_c' (C'- C'_0) + \alpha_t ( T-T_0)],
\end{equation}
where  $C$ is the concentration of soil particles and $C'$ is the concentration of solute impurities. 
The coefficients $\alpha_c$, $\alpha_c'$ and $\alpha_t$ are the particle, solutal and thermal expansion coefficients,
 respectively, taken to be constants.
We now let $C \to -C \Delta C + C_0, C' \to -C' \Delta C' + C'_0$ and $ T \to -T \Delta T + T_0 $. 
 We also define $B_c = \alpha_c \Delta C, B'_c = \alpha_c' \Delta C' $ and 
 $B_t = \alpha_t \Delta T $ and so we get, 
$
 \rho = \rho_0 [ 1 + B_c C + B_c' C' + B_t T ]
$.  
 Our system is described by a divergenceless vector field, which 
represents the fluid velocity, $
 \partial^i v_i =0
$.
In order to write the momentum balance equation, it is useful to define a   substantive derivative as follows
\begin{equation}
 D v_i = \partial_t v_i + v^j \partial_j v_i, 
\end{equation}
where 
$
 \partial_t =  \partial/\partial t, \,\,  \partial_i =  \partial/\partial x^i,
$
and  $ \partial^i \partial_i = \partial^2 $. 
In continuum mechanics, this describes describes the time rate of change of some physical quantity 
 for a material element subjected to a space-and-time-dependent velocity field. 
Now we can write the momentum balance equation as 
\begin{equation}
 \rho_0 D v_i = -  \partial_i p +  \partial^j \mu [1- C/C_p]^{-2} ( \partial_i  v_j + \partial_j  v_i) - \rho g \lambda_i.
\end{equation}
Here  $\rho$ is the density, $g$ is the acceleration 
due to gravity,  $\mu$ is a constant viscosity and 
$C_p$ is a maximum close packing concentration (shrinkage limit) \cite{wb1}. 
 The factor $[1- C/C_p]^{-2}$ denotes the dependence of viscosity on concentration. 
 Along with this equation there are also the following equations, 
\begin{eqnarray}
 DT &=& \kappa \partial^2  T, \nonumber \\ 
 DC &=& \partial^i d_{11} [1- C/C_p]^{-2}\partial_i C +  d_{12} \partial^2 C', \nonumber \\ 
 DC' &=&  d_{21}\partial^2 C +  d_{22} \partial^2 C',
\end{eqnarray}
where $\kappa$ is the thermal diffusivity of the fluid, $d_{11}$  is the soil diffusion coefficient,
 $d_{22}$ is the solute diffusion coefficient, and $d_{21}, d_{12}$ are  cross diffusion coefficients. 
We have  neglected both the Dufour effect and Soret effect  in the temperature equation  as they are very weak relative to 
other effects. Here the thermal diffusivity of the fluid, solute diffusion coefficient  and 
the cross diffusion coefficients are constants. 
As we will be analysing the this system far away from $C_p$, we will neglect the factor $C/C_p$ and  so 
in this limit  even soil diffusion coefficient and the viscosity is a constant. 
We will also use the Boussinesq approximation and set density constant in all terms except $\rho g \lambda_i$. 
We can use the natural length scale $l$, which is the  depth of the liquid, 
 and thermal diffusivity $k$ to non-dimensional these equations. 
To do so we will transform each quantity as $x^i \to l x^i,  \partial_i \to l^{-1} \partial_i,  v_i \to k l^{-1} v_i,
t \to l^2 k^{-1} t,$ and $p \to p \mu k l^{-2} + \rho_0 g (l- \lambda^i r_i) $. 
The new  velocity field again remains divergenceless,
$
 \partial^i v_i = 0
$.
The non-dimensional form of the momentum conservation equation and the continuity equation for temperature 
and concentration for constant viscosity and  diffusion coefficients becomes
\begin{eqnarray}
{Pr^{-1} D v_i  } &=& - \partial_ i p + \partial^2 v_i+ R_c C  \lambda_i + R_c' C'  \lambda_i +R_t T  \lambda_i , \nonumber \\ 
{D C }&=& Le^{-1}_{11} \partial^2 C  + Le^{-1}_{12} \partial^2 C', \nonumber \\ 
{D C' }&=& Le^{-1}_{21} \partial^2 C'  + Le^{-1}_{22} \partial^2 C' , \nonumber \\ 
{D T} &=& \partial^2 T, 
\end{eqnarray}
where $ Pr = k \rho_0 \mu^{-1}$ is the Prandtl number, $R_c = \mu^{-1} k^{-1}B_c \rho_0 g l^3 $ and 
 $R_c' = \mu^{-1} k^{-1}B_c' \rho_0 g l^3 $
are the concentration 
 Rayleigh numbers,  $R_t = \mu^{-1} k^{-1}B_t \rho_0 g l^3 $ is the temperature  
 Rayleigh number,  $Le_{11}  = k d^{-1}_{11 }, Le_{22} = k d^{-1}_{22} $ are the Lewis numbers and 
$Le_{12} = k d^{-1}_{12}, Le_{21} = k d^{-1}_{21}$ are new Lewis number corresponding to cross diffusivity. 
We want to analyze the steady state version of these equations. In order to do so, we impose the following 
boundary conditions, $\lambda^i v_i = 0 $, $  
\lambda^i \Lambda^j \partial_i v_j 
= 0 $, where $ 
\Lambda^i \lambda_i = 0$, and $C  =C'= T = 1$, $C(1)= C'(1)= T(1) =0$, $ p(1) =0$.  
We also take $\lambda_i = (0, 0, 1)$ and $r_i = (x, y, z)$, thus our liquid is placed on a level plane. 
Under these boundary conditions we can obtain the  steady state solutions. 
We will also add small perturbations to these steady state solutions. Now the steady  state solutions
with perturbations added to them is are given by 
\begin{eqnarray}
 v_i &=& u_i \nonumber \\
C &=& 1-\lambda^i r_i + \theta_c \nonumber \\
C' &=& 1-\lambda^i r_i + \theta_c' \nonumber \\
T &=& 1-\lambda^i r_i + \theta_t \nonumber \\
p &=&  -\frac{ (R_c + R_c'+ R_t)}{2}|1-\lambda^i r_i|^2  + \phi.
\end{eqnarray}                                                                 
Now we define $D^a_b = a^{-1} \partial_t - b^{-1} \partial^2$, and so we get 
\begin{eqnarray}
 D^{Pr}_1 u_i + \partial_i \phi - ( R_c \theta_c  +R_c' \theta_c ' +  R_t \theta_t )  \lambda_i&=&0, \nonumber \\
D^1_{Le11} \theta_c - Le_{12}^{-1} \partial^2\theta_c' - \lambda^i u_i &=&0, \nonumber \\
D^1_{Le22} \theta_c' - Le_{21}^{-1} \partial^2\theta_c - \lambda^i u_i &=&0, \nonumber \\
D^1_1 \theta_t - \lambda^i u_i &=&0.
\end{eqnarray}
We define an operator $E$ which takes the curl of any vector field $a^i$  twice, 
\begin{eqnarray}
  E a^i &=&\epsilon^{imn}\epsilon_{nkl} \partial_m \partial^k a^l \nonumber \\ 
&=&  (\delta^i_{k} \delta_{l}^m 
- \delta^i_{l}\delta_{k}^m )\partial_m \partial^k  a^l \nonumber \\ &=& 
\partial^i \partial^p a_p  - \partial^2 a^i.
\end{eqnarray}
So,   the gradient of a scalar field  vanishes when  its curl is taken twice, $ E  \partial^i \phi = 0 $,  
because in the expression 
$ E  \partial^i \phi = \epsilon^{imn}\epsilon_{nkl} \partial_m \partial^k \partial^l \phi $ there is a 
 contraction between a pair of symmetric and antisymmetric 
tensor indices. We also have 
$E  u_i = - \partial^2 u_i
$, because $u_i$ is divergenceless, and  $
\lambda^i (E  \lambda_i \phi) = \lambda^i\partial_i \lambda^i \partial_j \phi - \partial^2 \phi  = \tilde \partial^2 \phi
$. Now acting on
the momentum balance by $E$ and contracting it with $\lambda_i$,  we get 
\begin{eqnarray}
 D^{Pr}_1 \partial^2 \lambda^i u_i  - R_c \tilde \partial^2   \theta_c - R_c' \tilde \partial^2   \theta_c'
- R_t \tilde \partial^2   \theta_t = 0.
\end{eqnarray}
Now we  define the following differential operator 
\begin{eqnarray}
 \mathcal{D}_{0} &=& D^1_{Le 22} D^1_{Le 11} - Le^{-1}_{12}Le^{-1}_{21} \partial^4, \nonumber \\ 
 \mathcal{D}_{1} &=& (D^1_{Le22 } - Le^{-1}_{12} \partial^2)  , \nonumber \\ 
 \mathcal{D}_{2} &=& (D^1_{Le11} - Le^{-1}_{21} \partial^2)  , \nonumber \\ 
\end{eqnarray}
and the following Green's function's 
\begin{eqnarray}
  \mathcal{D}_{0} G_c (r, r' ) &=& \delta^{3} (r-r'), \nonumber \\
D^1_1 G_t (r, r') &=& \delta^{3} (r-r').
\end{eqnarray}
Now we have 
\begin{eqnarray}
   \mathcal{D}_{0} \theta_c &=& \mathcal{D}_{1}  \lambda^i u_i, \nonumber \\ 
   \mathcal{D}_{0} \theta_c' &=&  \mathcal{D}_{2} \lambda^i u_i, \nonumber \\ 
D^1_1 \theta_t&=&\lambda^i u_i.
\end{eqnarray}
So, we have 
\begin{eqnarray}
 \theta_c (r) &=& \int d^3 r' G_c (r, r' )  \mathcal{D}_{1}  \lambda^i u_i (r'), \nonumber \\ 
 \theta_c' (r) &=& \int d^3 r' G_c (r, r' )  \mathcal{D}_{2}  \lambda^i u_i (r'), \nonumber \\ 
 \theta_t (r)&=&\int d^3 r' G_t (r, r' )\lambda^i u_i (r).
\end{eqnarray}
because 
\begin{eqnarray}
\mathcal{D}_{0}   \theta_c (r) &=&\int dr' \mathcal{D}_{0}G_c (r, r')\mathcal{D}_{1} \lambda^i u_i (r'), \nonumber \\ 
 &=& \int d^3 r' \delta^3(r-r')\mathcal{D}_{1} \lambda^i u_i (r') \nonumber \\
&=& \mathcal{D}_{1}\lambda^i u_i (r), \nonumber \\ 
\mathcal{D}_{0}   \theta_c (r) &=&\int dr'\mathcal{D}_{0}   G_c (r, r')\mathcal{D}_{2}\lambda^i u_i (r'), \nonumber \\ 
 &=& \int d^3 r' \delta^3(r-r')\mathcal{D}_{2} \lambda^i u_i (r') \nonumber \\
&=& \mathcal{D}_{2}\lambda^i u_i (r), \nonumber \\ 
{  D^1_1  }\theta_t (r) &=&\int dr' D^1_1 G_t(r, r')\lambda^i u_i (r') \nonumber \\
&=& \int d^3 r' \delta^3(r-r')\lambda^i u_i (r') \nonumber \\
&=& \lambda^i u_i (r).
\end{eqnarray}
Thus, we can write, 
\begin{eqnarray}
 D^{Pr}_1 \partial^2 \lambda^i u_i (r) &=&  R_c \tilde \partial^2   \int d^3 r' G_c (r, r' )  \mathcal{D}_{1}  \lambda^i u_i (r')
\nonumber \\ && + R_c' \tilde \partial^2   \int d^3 r' G_c (r, r' )  \mathcal{D}_{2}  \lambda^i u_i (r') \nonumber \\ &&
+ R_t \tilde \partial^2   \int d^3 r' G_t (r, r' )\lambda^i u_i (r').
\end{eqnarray}
Multiplying by $\mathcal{D}_0$ and $D^1_1$, we get 
\begin{eqnarray}
 \mathcal{D}_0 D^1_1 D^{Pr}_1 \partial^2 \lambda^i u_i  &=& 
 R_c \tilde \partial^2  D^1_1 \mathcal{D}_{1}  \lambda^i u_i 
 + R_c' \tilde \partial^2   D^1_1  \mathcal{D}_{2}  \lambda^i u_i  \nonumber \\ &&
+ R_t \tilde \partial^2  \mathcal{D}_0 \lambda^i u_i .
\end{eqnarray}

Now using  the boundary conditions that the even derivatives of $\lambda^i \partial_i$
vanishes on $\lambda^i u_i $, we  write the solution as \cite{ti}, 
\begin{equation}
 \lambda^i u_i = A \sin n\pi \, exp [i(k_x x + k_y) + \sigma t],
\end{equation} becomes
and  we also define 
\begin{eqnarray}
 k^2 = k_x^2 + k^2_y, && k_n^2 = k^2 + n^2 \pi^2. 
\end{eqnarray}
Thus the { characteristic} equation is 
\begin{equation}
 C(k_m, k, \sigma ) =0, 
\end{equation}
where 
\begin{eqnarray}
 C(k_m, k, \sigma )& =& (  (\sigma + k_n^2 Le^{-1}_{22}  )  (\sigma + k_n^2 Le^{-1}_{11}  ) - Le^{-1}_{12}Le^{-1}_{21} k_n^4 )\nonumber \\ && \times 
 (\sigma + k_n^2  )  (\sigma Pr^{-1} + k_n^2  ) k_n^2 \nonumber \\ 
&&- R_c (\sigma + k_n^2   )  ((\sigma + k_n^2 Le^{-1}_{22}  )  - Le^{-1}_{12} k_n^2) k^2 \nonumber \\ 
&& - R_c' (\sigma + k_n^2   )  ((\sigma + k_n^2 Le^{-1}_{11} )  - Le^{-1}_{12} k_n^2) k^2\nonumber \\ 
&& - R_t (  (\sigma + k_n^2 Le^{-1}_{22}  )  (\sigma + k_n^2 Le^{-1}_{11}  ) \nonumber \\ && - Le^{-1}_{12}Le^{-1}_{21} k_n^4 ) k^2.
\end{eqnarray}
 Assuming the principle of exchange of stabilities  critical behavior is 
obtained by setting $\sigma =0$ \cite{ti},  we have 
\begin{equation}
  C(k_m, k, 0 ) =0, 
\end{equation}
where 
\begin{eqnarray}
 C(k_m, k, 0)&=& (   Le^{-1}_{22}    Le^{-1}_{11}   - Le^{-1}_{12}Le^{-1}_{21} )
 k_n^6 k^{-2} \nonumber \\ 
&&- R_c      (  Le^{-1}_{22}    - Le^{-1}_{12} ) \nonumber \\ 
&& - R_c'     (  Le^{-1}_{11}   - Le^{-1}_{12} ) \nonumber \\ 
&& - R_t (    Le^{-1}_{22}   Le^{-1}_{11}   - Le^{-1}_{12}Le^{-1}_{21} ).
\end{eqnarray}
 Now we can define a effective  Rayleigh number as 
\begin{eqnarray}
 R &=& [R_c      (  Le^{-1}_{22}    - Le^{-1}_{12} )+ R_c'     (  Le^{-1}_{11}  
 - Le^{-1}_{12} )\nonumber \\ &&+ R_t (    Le^{-1}_{22}   Le^{-1}_{11}   - Le^{-1}_{12}Le^{-1}_{21} ) ]\nonumber \\ &&\times 
(   Le^{-1}_{22}    Le^{-1}_{11}   - Le^{-1}_{12}Le^{-1}_{21} )^{-1}.
\end{eqnarray}
Hence, we can write $R = k_n^6 k^{-2}$.
The lowest value is for $n =1$ and the instability starts at 
\begin{equation}
 \frac{\partial  R}{\partial k^2} = 0.
\end{equation}
This gives us 
$
 k^2 = \pi^2/2
$, and 
the corresponding value of $ R$ will  be given by 
$ R= 657.5  \sim 10^3.
$

This is when the instability will start. 
This convection in  ternary mixtures can be used to analyse
interesting geological phenomenon.  This is because soils can exhibit semi-permeability and 
osmosis through these semipermeable
soils can be driven by  gradient in salt concentration \cite{os1, os2, os0}.  
In fact, an  excess pressure   has been observed to exist  during the 
diffusion of salty water through certain soils  \cite{os4, os5}.
Thus, it will be interesting to analyse this phenomenon using 
 stability analysis for ternary systems. In the limit when there is no cross diffusion, $Le^{-1}_{12} = Le^{-1}_{21} =0$, 
we have the expected result $R = R_c L_{11} + R_c' L_{22} + R_t$. Thus,  in this limit, the temperature 
Rayleigh number and all the concentration 
Rayleigh multiplied by there respective Lewis numbers add up to give the effective Rayleigh number. 
So, the effective Rayleigh number is enhanced by the existence of a salt gradient. 
This, also implies that convection 
can start much earlier in salty water than pure water. Thus, 
the evaporation of salty water at the surface of  soils 
can set up interesting instabilities, governed by the ternary characteristic equation.

\section{Binary Mixture}
In this section we will neglect the effect due to impurities. However, we are not able to directly set $Le^{-1}_{22} = Le^{-1}_{21} = Le^{-1}_{12} =0$, 
in the effective Rayleigh number. This is because the Green's function used to derive this effective Rayleigh number is not well defined for these 
values. This is because the differential operator $\mathcal{D}_0$ has a zero eigenvalue for  these values of the Lewis numbers. So, its 
inverse does not exist. Thus, we can not set $Le^{-1}_{22} = Le^{-1}_{21} = Le^{-1}_{12} =0$, in the effective Rayleigh number
obtained by using this Green's function. 
To obtain a correct result we     will have to repeat the above analysis for the binary solutions. However, it may be noted that 
all the analysis of the previous section, except the derivation of the Green's function remains, 
remains  well defined,  if we set $Le^{-1}_{22} = Le^{-1}_{21} = Le^{-1}_{12} =0$. So, now 
we set, $Le_{11} = Le,  Le^{-1}_{22} = Le^{-1}_{21} = Le^{-1}_{12} =0, \rho = \rho_0 [1 + B_c C + B_t T]$, and use the following equations 
\begin{eqnarray}
{Pr^{-1} D v_i  } &=& - \partial_ i p + \partial^2 v_i+ R_c C  \lambda_i + R_t T  \lambda_i , \nonumber \\ 
{D C }&=& Le^{-1} \partial^2 C  , \nonumber \\ 
{D T} &=& \partial^2 T.
\end{eqnarray}
If we repeat the above analysis, 
we will obtain the following perturbative equations  
\begin{eqnarray}
 D^{Pr} \partial^2 \lambda^iu_i - R_c \tilde \partial^2 \theta_c - R_t \tilde \partial^2 \theta_t &=&0,  \nonumber \\
D^1_{Le} \theta_c - \lambda^iu_i &=& 0,  \nonumber \\
D^1_1 \theta_t - \lambda^iu_i &=& 0. \nonumber \\
\end{eqnarray}
Now we again define the following   Green's functions 
\begin{eqnarray}
 D^1_{Le} G (r, r')&=&  \delta^3 (r-r'), \nonumber \\ 
 D^1_{1} G_t (r, r')&=&  \delta^3 (r-r').
\end{eqnarray}
Thus, we can write 
\begin{eqnarray}
 \theta_c (r) &=& \int d^3 r' G (r,r') \lambda^i u_i (r'), \nonumber \\ 
 \theta_t (r) &=&  \int d^3 r' G_t (r,r') \lambda^i u_i (r'),
\end{eqnarray}
because 
\begin{eqnarray}
D^1_{Le} \theta_c (r) &=& \int d^3 r'D^1_{Le} G (r,r') \lambda^i u_i (r') \nonumber \\ 
 &=& \int d^3 r' \delta (r-r') \lambda^i u_i (r') \nonumber \\ 
&=& \lambda^i u_i (r), \nonumber \\ 
D^1_{1} \theta_t(r) &=& \int d^3 r'D^1_{1} G_t (r,r') \lambda^i u_i (r') \nonumber \\ 
 &=& \int d^3 r' \delta (r-r') \lambda^i u_i (r') \nonumber \\ 
&=& \lambda^i u_i (r).
\end{eqnarray}
We can write 
\begin{eqnarray}
  D^{Pr} \partial^2 \lambda^iu_i (r) &=& R_c \tilde \partial^2 \int d^3 r' G(r,r') \lambda^i u_i (r') \nonumber \\ 
&& +  R_t \tilde \partial^2 \int d^3 r' G_t (r,r') \lambda^i u_i (r'). 
\end{eqnarray}
Now acting on this equation by $D^1_1$ and $D^1_{Le}$, we get 
\begin{equation}
 D^1_1 D^1_{Le} D^{Pr}_1 \partial^2 \lambda^iu_i = R_c \tilde \partial^2 D^1_1 \lambda^i u_i  
 + R_t \tilde \partial^2 D^1_{Le} \lambda^i u_i (r').
\end{equation}
So, again using  the boundary conditions that the even derivatives of $\lambda^i \partial_i$
vanishes on $\lambda^i u_i $, we  write the solution as \cite{ti}
\begin{equation}
 \lambda^i u_i = A \sin n\pi \, exp [i(k_x x + k_y y) + \sigma t],
\end{equation}
Thus, we get  
\begin{equation}
 C(k_n, k, \sigma) =0, 
\end{equation}
where 
\begin{eqnarray}
 C(k_n, k, \sigma) &=& (\sigma + k_n^2 Le^{-1}  ) (Pr^{-1}\sigma + k_n^2  ) (\sigma + k_n^2   ) k_n^2 \nonumber \\ & &
- (\sigma + k_n^2 Le^{-1}  ) k^2 R_t  - 
(\sigma + k_n^2  ) k^2 R_c. 
\end{eqnarray}
So, in the case of temperature dependence critical behavior is 
obtained by setting $\sigma =0$,  we have \cite{ti}
\begin{equation}
  C(k_m, k, 0 ) =0, 
\end{equation}
where 
\begin{eqnarray}
 C(k_n, k, 0) &=&  k_n^6 k^{-2}Le^{-1}  
-  Le^{-1}   R_t  -  R_c. 
\end{eqnarray}
Now here the effective   Rayleigh number is
\begin{equation}
 R = R_t + Le R_c
\end{equation}
Hence, we can write $
 R = k_n^6 k^{-2}.
$
The lowest value is for $n =1$ and the instability starts at 
\begin{equation}
 \frac{\partial R}{\partial k^2} = 0.
\end{equation}
This gives us 
$
 k^2 = \pi^2/2
$, and 
the corresponding value of $ R$ will be again be given by 
$
 R= 657.5  \sim 10^3
$.

Now the actual value for a binary mixture of  bentonite and water are of the order,  $
\rho_0 \sim 10^3, \mu \sim 10^{-3},  \kappa \sim 10^{-7}$. Also the depth of the liquid 
$ l \sim 10^{-2}$  \cite{ee} and $B_c \sim 1, B_t \sim 10^{-1}$ \cite{ee11}, so we have
 $R_t \sim 10^3$, and $R_c \sim 10^6$. However,  in the stability analysis the product 
of $R_c$ and $Le$  contributes to the occurrence of the instability and $Le \sim 10^2$. So, the contribution from the concentration gradient 
$R_c Le \sim 10^8$ is much more than temperature gradient. The   Marangoni effect is also measured by the 
Marangoni number $M\sim 10^3$ \cite{ee22}. Hence, the concentration gradient is the most dominant factor for convection to occur. 
The calculated value  of the effective Rayleigh number due to the
concentration gradient is much larger than the theoretical limit for
the  convection to occur. Hence, we hypothesize that it is responsible for the  instability in the drying of a binary mixture of
 bentonite and water. 
 
It has been observed that in a binary mixture of  bentonite and water initially the convection takes place on the surface, but, 
  eventually a  layer forms on the surface and this inhibited further convection from taking place on the surface  \cite{ee}. 
However, the convection continues in the lower depths of the soil. This observation rules out the possibility that 
this convection is due to the Marangoni effect. This is because in that case 
 it would be surface driven phenomenon and would not continue at depth after halting at the surface.
It will be interesting to explore the possibility that this behavior is due to the fact that in a soil the diffusivity and
 viscosity are strong functions of particle concentration. That is, the effective viscosity and diffusivity of a soil slurry 
increase dramatically with increasing particle concentration, and thus at sufficiently high particle concentrations the effective 
Rayleigh number will be reduced to below the critical value  necessary for convection.

\section{Conclusion}
In this paper we have analysed the drying of a ternary  mixture of  soil particles and water mixed with impurities.
We have demonstrated that 
the concentration gradient that forms during the process of evaporation of this mixture is the main driving force  for the 
occurrence of a Rayleigh-Benard instability. This instability causes convection to take place, which in turn 
 may have significant implications for the drying of soil crusts and the formation of patterned ground.
 In the ternary case the use of Green's functions allowed for a straight-forward calculation of the characteristic equation.
 For the binary case a separate calculation was required. This is because the differential operator that was inverted using the 
 Green's function in the ternary case, yields a zero eigenvalue   in the binary case. Thus, it cannot be inverted and the ternary Green's function 
becomes ill defined for the binary case. 
 As noted in the introduction, soil loss in agricultural areas depends critically 
on the formation of surface seals and crusts and the subsequent cracking of these crusts during  drying of the ground. 
It is possible that the convection patterns give rise to weak points, which on drying gives rise to cracks \cite{ee}. 
Furthermore, patterns observed during the freezing and thawing of ground around the Arctic circle have also been explained
 as owing to convection of the soil. 
In this paper it was proposed that evaporation from the surface of thawing ground causes an unstable density 
profile to develop potentially leading to convection. The occurrence of similar patterns on Mars suggests that in the past
 the temperature on Mars may have been large enough for the ice to melt and evaporation to take place. 
It may be noted that it has been observed that drying on slanting slopes gives rise to rolls and drying on flat 
surface gives rise to hexagonal structures \cite{ara0}. 
It will be interesting to perform stability analysis for these specific physical situations and see if the formation of these 
particular patterns can be explained.

\end{document}